\documentclass[aps,prl,reprint,superscriptaddress,amsfonts,amssymb,amsmath,showpacs]{revtex4-1} 

\usepackage{amsmath,amssymb}
\usepackage[T1]{fontenc}
\usepackage[utf8]{inputenc} 
\usepackage{graphicx}
\usepackage[dvips]{epsfig}
\usepackage{xcolor}
\usepackage{color}
\usepackage{float}
\usepackage{lipsum}
\usepackage{microtype}
\usepackage{siunitx}
\usepackage{ulem}
\usepackage[colorlinks=true, linkcolor=royal, urlcolor=royal, citecolor=royal, anchorcolor=royal]{hyperref}

\definecolor{royal}{RGB}{0,0,50}
\definecolor{greenSimon}{RGB}{20,150,40}

\usepackage[colorinlistoftodos]{todonotes}

\begin{document}

% PREVIOUS TITLES
% Mass transport measurement through individual nanochannel using flow-induced concentration polarization.
% Flow-induced concentration polarization for ultra-sensitive mass transport measurement through a single nanochannel.
% Flow-induced concentration depletion for ultra-sensitive mass transport measurement through a single nanochannel.
\title{Flow-Induced Shift of the Donnan Equilibrium for Ultra-Sensitive Mass Transport Measurement Through a Single Nanochannel.}

\date{\today}

\author{Simon Gravelle}
\affiliation{Univ Lyon,Universit\'e Claude Bernard Lyon1, CNRS, Institut Lumi\`ere Mati\`ere, F-69622 Villeurbanne, France}
\affiliation{School of Engineering and Material Science, Queen Mary University of London, UK}
\author{Christophe Ybert}
\affiliation{Univ Lyon,Universit\'e Claude Bernard Lyon1, CNRS, Institut Lumi\`ere Mati\`ere, F-69622 Villeurbanne, France}

\begin{abstract}
Despite mass flow is arguably the most elementary transport associated to nanofluidics, its measurement still constitutes a significant bottleneck for the development of this promising field. Here, we investigate how a liquid flow perturbs the ubiquitous enrichment --or depletion-- of a solute \textit{inside} a single nanochannel. Using Fluorescence Correlation Spectroscopy to access the local solute concentration, we demonstrate that the initial enrichment --the so-called Donnan equilibrium-- is depleted under flow thus revealing the underlying mass transport. Combining theoretical and numerical calculations beyond the classical 1D treatments of nanochannels, we rationalize quantitatively our observations and demonstrate unprecedented flow rate sensitivity. Because the present mass transport investigations are based on generic effects, we believe they  can develop into a versatile approach for nanofluidics. 
\end{abstract}

\maketitle

%%%%%%%%%%%%%%%%%%%%%%%%%%%
% Introduction 
%%%%%%%%%%%%%%%%%%%%%%%%%%%

Nanofluidics, the fluidic transport in nanoscale confined systems, is encountered in a wide range of situations spanning from biological systems to filtration and colloidal science \cite{Spiegler1966,Gross1968}. Despite its ubiquity and longlasting history, it has experienced a strong and fast growing interest over the past few years, thanks to our increasing ability to fabricate and characterize nanoscale systems \cite{Bocquet:2010ui,Bocquet:2014gu,Segerink:2014hz}. This led notably to the identification of striking phenomena, among which giant permeability enhancement in carbon nanotubes \cite{Holt:2006vr,Falk:2010cr,Secchi:2016bb}, original energy transduction pathways \cite{Pennathur:2007ti,VanDerHeyden:2007vo,Siria:2013dx} or complex electrokinetic transports \cite{Daiguji2005, Karnik2007, Deng2013}.

At the roots of these phenomena usually lies one major property of nanofluidics that originates from the enhanced surface effects: the intimate interplay between transports and drivings of different nature: electrical, hydrodynamic, solutal or thermal. So far however, only ionic transport is routinely captured at the scale of a single nanochannel \cite{Branton:2008fr}, while disentanglement of the physical mechanisms at stake would require independent access to the different transport currents. Surprisingly, the flow rate measurement at the scale of single nanofluidic object -- arguably the most basic fluidic transport  -- still constitutes a serious bottleneck for quantitative studies.

%%%%%%%%%%%%%%
%
%
\begin{figure}[h!]
\includegraphics[width=\linewidth]{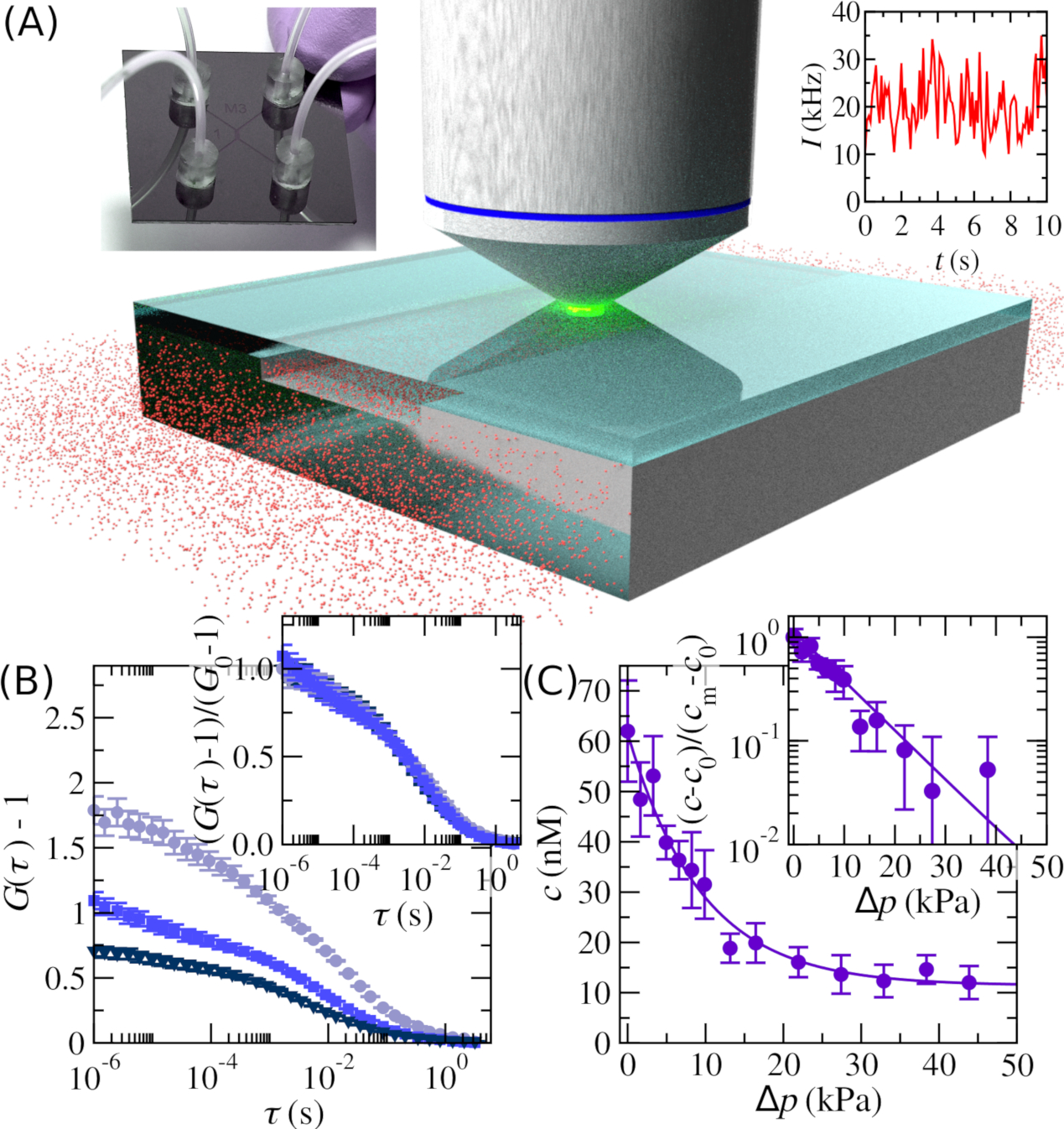}
\caption{%
\textbf{(A)} Experimental setup: nanofluidic silicon chip (left); fluorescence detection configuration (middle); raw fluorescence-intensity fluctuations $I(t)$ (right). \textbf{(B)} Fluorescence intensity auto-correlation function $G(\tau)-1$ under different pressure-driven flows ($\Delta p=0$, $6$ and $15$\,kPa, respectively $\text{Pe}=0$, $30$ and $70$). Inset: same data normalized by the value at vanishing lagtime $(\tau=10^{-6}\,\text{s})$. \textbf{(C)} Measured dye concentration \textit{vs} pressure differences ($L = 25\,$\si{\micro}m; $c_s=10\,$mM). Inset: same data normalized (see text) and in semi-log scale.
}
\label{fig:FIGURE1}
\end{figure}
%
%
%%%%%%%%%%%%%%

In order to meet the sensitivity required for single nanochannel flow rate measurements, significant efforts have been made over the past few years. This includes electrical measurement approaches such as cross-correlation for nanochannels \cite{Mathwig2012} or Coulter counting for transmembrane nanopores \cite{gadaleta2015ultra}. In addition to electrical approaches, important advances have been made using direct monitoring of volume change in reservoirs \cite{sharma2018direct} or optical velocimetry approaches. For pipette-like geometries, the peculiar properties of the Landau-Squire nanojet have been used to access exit flow rates via optical-tweezer based vorticity maps \cite{Laohakunakorn2013} and more recently by PIV flow maps \cite{Secchi:2016bb} or chemical plumes \cite{Secchi:2017ib}. Finally, for the benchmark geometry of nanochannels, a milestone in sensitivity, around a few fL/s, has been reached by some of the present authors by imposing fluorescent dye gradients along the system \cite{Lee2014c}. Despite such advances, no approach has currently emerged that could address the various needs in terms of experimental configuration or required sensitivity. Exploring alternative routes on mass transport signature in nanofluidic systems thus remains an important issue both from the fundamental point of view or from the practical perspective of developing the versatility of the nanoscale toolbox.

In this context, we consider here a generic property of solutes in confined systems, that are either enriched or depleted with respect to their bulk reservoir concentration. This enrichment (or depletion), called Donnan equilibrium, is due to specific solute--surface interactions whose characteristic scales are typically nanometric \cite{Bocquet:2010ui} and thus comparable with the confining scale. In the case of electrolytes under strong confinement, Debye layers overlap and only the counter-ion contributes to the ion concentration in the channel. This limit gives rise to a nanochannel ion-selectivity at the origin of striking nanofluidic properties \cite{Schoch2008}. 
%\sout{In the case of electrolytes, this gives rise for strong confinements, where Debye layers overlap, to nanochannel ion-selectivity, at the origin of striking nanofluidic properties \cite{Schoch2008}}. 
Among these, ion concentration polarization effects, that develop under an electric field forcing, promise huge practical applications such as water desalination \cite{Kim2010} and purification \cite{Deng2015}. Physically, such concentration gradients appear because conservation laws encompassed in the Nernst-Planck equation
\begin{equation}
\boldsymbol{\nabla} \cdot \left( - D \boldsymbol{\nabla} c - Z  c D \boldsymbol{\nabla} \phi + \boldsymbol{u}  c \right) = 0,
\label{diff_conv}
\end{equation}
where $D$ is the diffusion coefficient, $k_BT$ the thermal energy, $\phi = eV / (k_B T)$ the reduced electrical potential, $\boldsymbol{u}$ the velocity field and $Z$ the species valence, have to be enforced at the junction between media with different properties --the reservoirs and the ion-selective nanochannel.

In the following, we examine a complementary configuration where a nanochannel responds to a pressure forcing rather than an electric one. Using Fluctuation Correlation Spectroscopy (FCS) to access the local dye concentration \textit{within} a nanochannel, we show that convection induces a concentration depletion (or enrichment) that we refer to as a shift in the Donnan equilibrium. This concentration depletion/enrichment can be used to extract information about mass fluxes. 
%\sout{Using Fluctuation Correlation Spectroscopy (FCS) to access the local dye concentration \textit{within} a nanochannel, we show that it polarizes under flow with the dye concentration varying monotonically along the channel, and thus contains information about mass fluxes.} 
This effect holds even far from the Debye layer overlap regime, and we present theoretical and numerical analysis to go beyond the classical 1-dimensional (1D) approaches adapted to highly confined systems. Overall quantitative agreement is achieved between experiments and calculations from which ultra-small single nanochannel flow rates can be deduced. Based on generic %\sout{polarization} 
effects, we believe that both the physical mechanisms and the technique --highly sensitive and involving very low dye levels throughout the entire system-- can open perspectives for future investigations in the field of nanofluidics.

%%%%%%%%%%%%%%%%%%%%%%%%%%%%%%%%%%%
% Method and Experimental results %
%%%%%%%%%%%%%%%%%%%%%%%%%%%%%%%%%%%

Practically, measurements are performed inside slit nanochannels of height $h \simeq 150$\,nm, width $w = 5$\,\si{\micro}m and length $L=25$\,\si{\micro}m or $500$\,\si{\micro}m, engraved inside silicon chips as shown in Fig.\,\ref{fig:FIGURE1}\,A (see Supp. Mat. for fabrication details). %\sout{\cite{Lee2014c}} for fabrication details). 
The system is filled with a potassium chloride (KCl) solution of concentration $c_s$ ranging from $10^{-4}$ to 10$^{-2}$\,M supplemented with $c_0 = 10-100\,$nM of Rhodamine 6G fluorescent dye (R6G). Under such conditions, the silicon oxide nanochannel walls are expected to bear a negative charge as experimentally verified in similar systems \cite{Lee2014c}, with wall potential in the range $[-90, -50]$\,mV. Because R6G is a cationic dye, a surface excess develops close to solid walls leading to a global concentration enhancement in the confined nanochannel. A pressure forcing $\Delta p$ is imposed between the ends of the nanofluidic system using either a liquid column or a pressure controller (Elveflow). The generated flow is deduced from the Hagen–Poiseuille relationship $Q = \Delta p \, w h^3/(12 \eta L)$, with $\eta$ the water viscosity, with a mean flow velocity $\langle u\rangle = Q / (h w)$.

The nanofluidic chip is located onto the stage of an inverted microscope (Nikon TE-2000U) with a water immersion objective (60x, NA=1.2). A parallel laser beam (CNI, 532\,nm) filling the back pupil of the objective generates a diffraction-limited spot in the sample (diameter $d \sim 500$\,nm). A confocal excitation and detection pathway is used to collect the dye fluorescence intensity signal $I(t)$, with the signal filtered through a confocal pinhole to reject out-of-focus contributions before being sent towards two avalanche photo-diode detectors (Perkin Elmer), see raw signal in Fig.\,\ref{fig:FIGURE1}\,A. The collected raw intensity is analyzed through the autocorrelation function $G(\tau)$ which contains information about the probe concentration and dynamics \cite{Magde1974, Digman2011, Koppel1974} (Fig.\,\ref{fig:FIGURE1}\,B). 

Classical FCS velocimetry would rely on analyzing characteristic decay times in $G(\tau)$ \cite{Palmer:1987uk,Yordanov:2009ui}. In the present context however, such approach is meaningless as single point FCS velocimetry has a "poor" sensitivity threshold set by the --small-- focal spot size $\langle u\rangle > D/d \sim 1$\,mm/s. Accordingly, normalized autocorrelations obtained with or without pressure-driven flows superimpose perfectly showing no evolution in time scales, as represented in the inset of Fig.\,\ref{fig:FIGURE1}\,B. Note that this master shape is rendered complex by the existence of reversible adsorption that gives rise to a slow time scale $\sim 10$\,ms atop the fast free-diffusion one $\sim 10$\,\si{\micro}s \cite{Durand2009, Daniels2010, Fowlkes2013b, Patel2013a} (see Supp. Mat. for further discussion). 

Despite the flow having no measurable impact on time scales, it has a strong and straightforward effect on the measured autocorrelation functions inside the nanochannel, as can be seen in Fig.\,\ref{fig:FIGURE1}\,B: the stronger the flow, the larger the amplitude of $G(\tau)$. This effect is a signature of a flow-induced decrease of the dye concentration at the --fixed-- measurement location within the nanochannel \cite{Krichevsky:2002vs}. Indeed the local concentration in the probed volume is known to obey $\langle c \rangle\propto [G(\tau\rightarrow0)-1]^{-1}$.

Fig.\,\ref{fig:FIGURE1}\,C shows how the measured concentration at the central location in the nanochannel evolves as a function of the imposed pressure differences $\Delta p$, for a nanochannel of length $L=25\,$\si{\micro}m and a salt concentration of $c_s=10\,$mM. Starting from a reference concentration under no flow, the solute concentration decreases with $\Delta p$ until it reaches a constant value. Qualitatively, under quiescent conditions the nanochannel has a higher concentration $c_m$ set by the equilibrium between the confining slit and the neighboring reservoirs at $c_0$. Under flow, the solute influx from reservoir is thus $J=c_0 Q$ while being $c_m Q>J$ inside the nanochannel. This mismatch will induce polarization effects to arise in the system and overall the nanochannel solute content will decrease under flow. Under strong flows, incoming solute concentration will set the one in the nanochannel hence to $c_0$.
More quantitatively, the inset of Fig.\,\ref{fig:FIGURE1}\,C presents the same concentration data, normalized according to $(c-c_0)/(c_m-c_0)$ and in semi-log scale. As can be clearly seen, the experimental concentration decreases exponentially toward the reservoir concentration $c_0$ upon pressure difference increase.

%
%%%%%%%%%%%%%%
%
%
\begin{figure}
\includegraphics[width=1.0\linewidth]{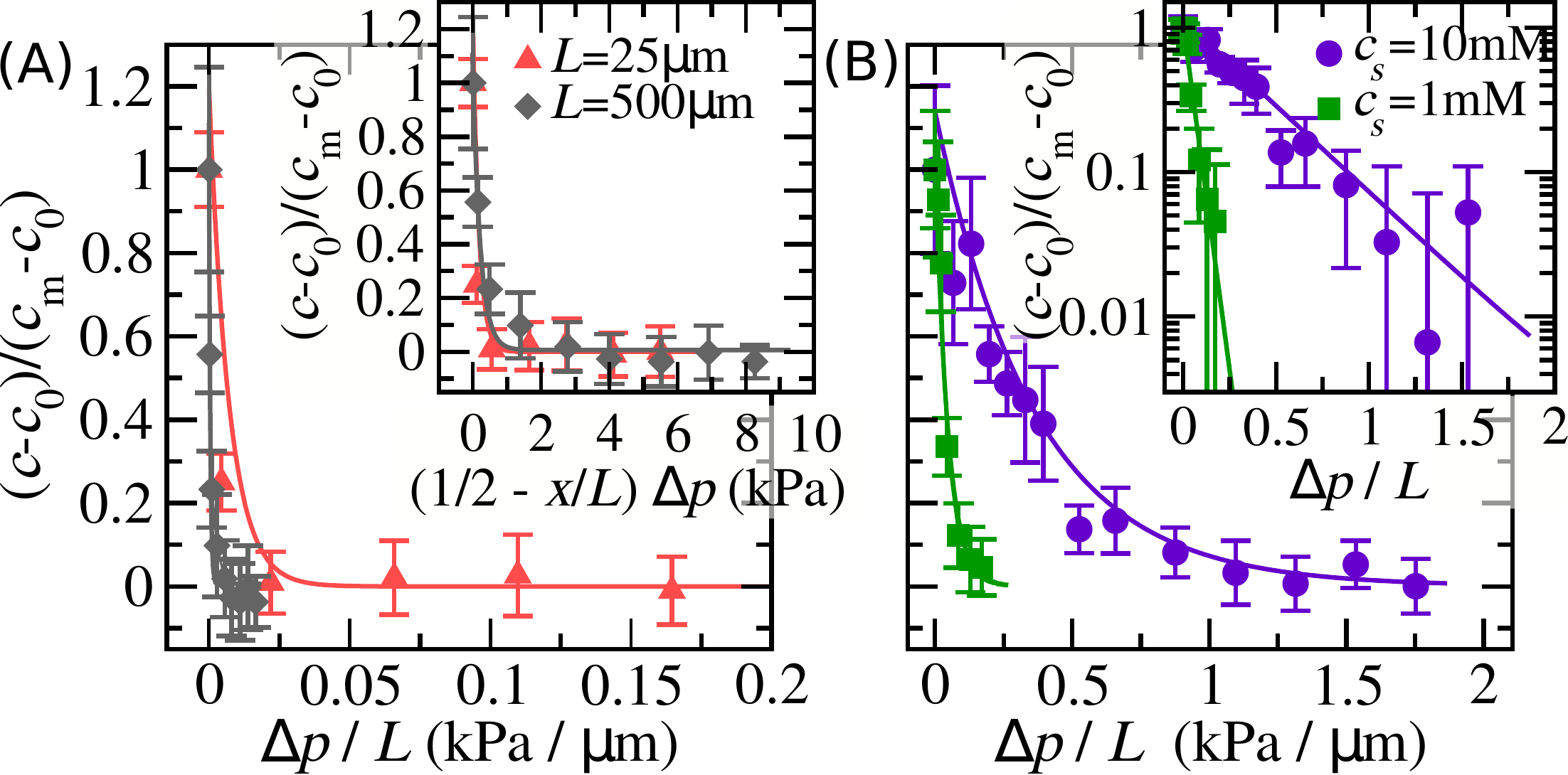}
\caption{%
\textbf{(A)} Normalized probe concentration as a function of pressure gradient $\Delta p / L$, for two nanochannel lengths and for $c_s=0.1\,$mM. ({\color{red}$\blacktriangle$}): $L=25$\,\si{\micro}m probed in its middle $x=0$; ({\color{gray}$\blacklozenge$}): $L=500$\,\si{\micro}m probed at its entrance $x=-L/2$. Inset: same curves as a function of pressure difference $(1/2 - x/L) \Delta p$, see text for the position-dependent scaling factor. The full lines: prediction from Eq. \eqref{eq:c3bis}. \textbf{(B)} Same as (A), for different salt concentrations and thus Debye layer overlap ratio $2 \lambda_D / h$, ($L=25$\,\si{\micro}m probed in their middle $x=0$). ({\color{greenSimon}$\blacksquare$}): $c_s = 1$\,mM salt concentration; ({\color{violet}$\bullet$}): $c_s=10$\,mM salt concentration. Inset: same data in semi-log scale.
}
\label{fig:FIGURE2}
\end{figure}
%
%
%%%%%%%%%%%%

Going further into the characterization, Fig.\,\ref{fig:FIGURE2} shows how these flow-induced effects depend on the experimental parameters. Changing the nanochannel length from $L=25\,$\si{\micro}m to $L=500\,$\si{\micro}m, we recover the same qualitative concentration decay (Fig.\,\ref{fig:FIGURE2}\,A). Quantitatively, the proper scale for concentration decay proves not to be the imposed pressure gradient $\Delta p/L$, which sets the flow velocity $\langle u\rangle$, but rather the absolute pressure difference $\Delta p$ (inset of Fig.\,\ref{fig:FIGURE2}\,A). This suggests the Peclet number \textit{over the channel length} to be the relevant variable $\mathrm{Pe}=\langle u\rangle L/D$ with $D=4 \cdot 10^{-6}$\,cm$^2$/s the dye diffusion coefficient, yielding a sensitivity orders of magnitude ($L/d$) higher than for the standard point-like FCS velocimetry. Note that 
as will be shown hereafter, the location $x \in [-L/2, L/2]$ where the concentration is probed enters as a normalizing factor $\alpha_{1D}=(1/2-x/L)$ of the Peclet. Now changing the salt background concentration at fixed $L$, the exponential concentration decay with $\Delta p$ is preserved but not the data collapse, with faster probe concentration decay occurring under smaller salt concentration (Fig.\,\ref{fig:FIGURE2}\,B). Such effect demonstrates that Peclet number alone is not sufficient to rationalize the full behavior of the nanosystems.

%%%%%%%%%%%%%%%%%%%%%%%%%
% One dimensional model %
%%%%%%%%%%%%%%%%%%%%%%%%%
%
%
%

To progress further in the understanding, we now develop a theoretical description of the problem. As a first step, we consider the classical assumption of a 1D situation where all physical quantities only depends on the $x$ position along the channel axis. The system considered is made of a nanochannel of length $L$ connected to two reservoirs of length $L_R \gg L$. Inside the nanochannel, we assume a homogeneous Donnan electric potential $V_D$, originating from the channel wall charges. From the reservoirs to the nanochannel, we consider that the electric potential linearly builds from zero to $V_D$ over thin transition regions of extent $\epsilon \ll L$, thus yielding a system made out of five different regions (see Supp. Mat. for details). Such 1D approach is in principle restricted to strong overlap conditions $2\lambda_D/h\gg1$, but is classically used beyond and has often proven effective \cite{Dydek2011}. The solute concentration $c_0$ is imposed at the far ends $x=\pm (L_R+\epsilon+L/2)$ of the reservoirs, and a flow velocity $u$ is imposed all along the system. Due to the difference in diffusivities -- about one order of magnitude \cite{Harned1947, Culbertson2002} --, we will neglect convection effects for the salt, with electric potential profiles thus keeping their at-equilibrium characteristics. Formally this corresponds to restricting to moderate Peclet numbers such that $\mathrm{Pe}\le D/D_s$ (see Supp. Mat. for numerical validation and details).

%%%%%%%%%%%%
%
%
\begin{figure}
\includegraphics[width=1.0\linewidth]{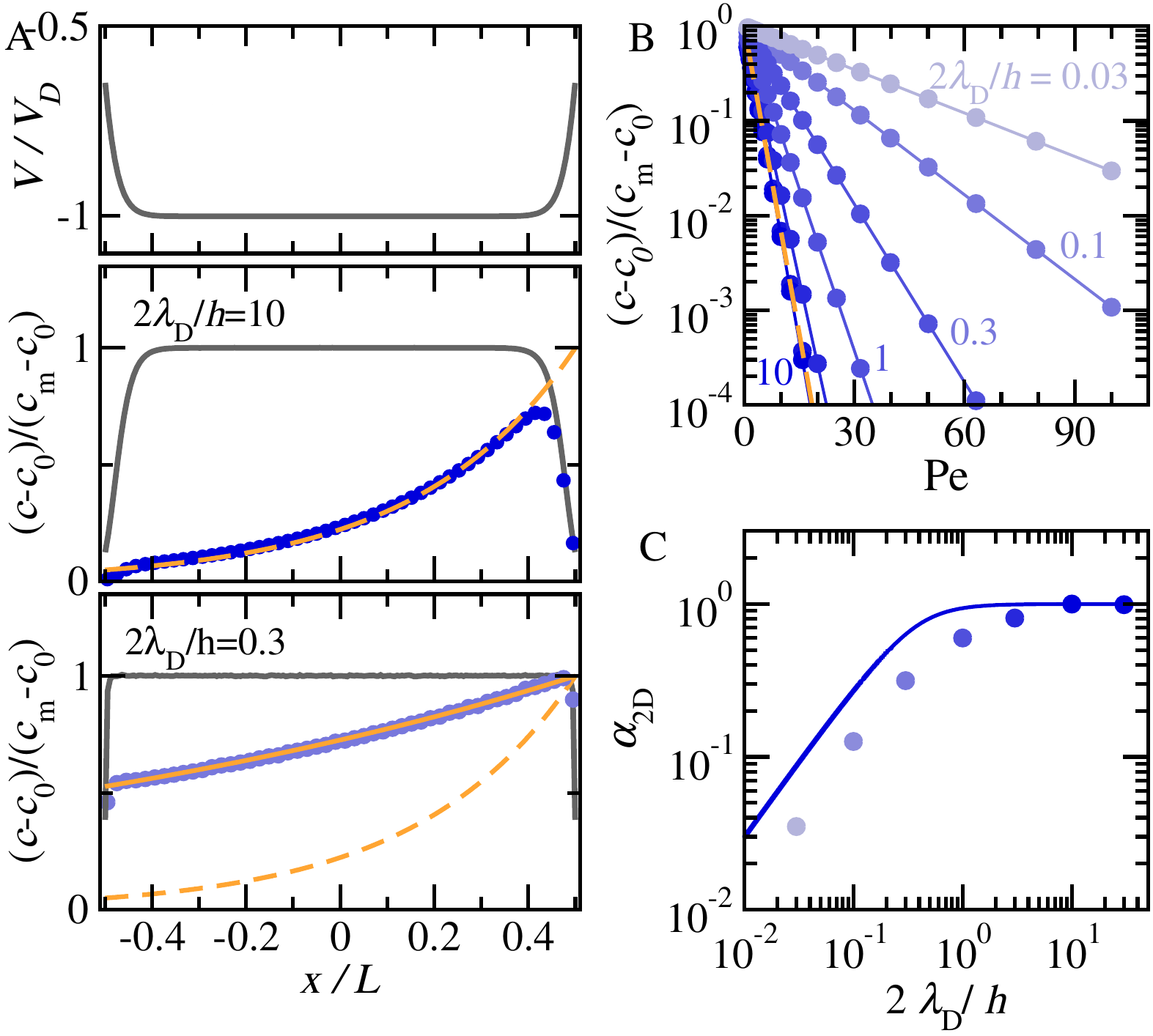}
\caption{%
Finite element calculations. 
\textbf{(A)} z-averaged potential (up) and dye concentration profiles within the channel (top and middle: $2\lambda_D/h=10$; low: $2\lambda_D/h=0.3$). ({\color{gray}---}): equilibrium profiles under no convection; blue symbols: profiles under convection ($\text{Pe}=3$). ({\color{orange}$--$}): 1D theoretical prediction Eq. (\ref{eq:c3bis}); ({\color{orange}---}): 2D theoretical prediction Eq. (\ref{eq:caveraged2}).
\textbf{(B)} z-averaged dye concentration at the middle location ($x=0$) as a function of the Peclet number for various confinements ($2 \lambda_D / h \approx 0.03$, $0.1$, $0.3$, $1$, $3$, $10$ and $30$). ({\color{orange}$--$}): same as in (A).
\textbf{(C)} Two-dimensional correction $\alpha_{2D}$ as a function of the confinement ratio $2 \lambda_D / h$. ({\color{blue}---}): analytical prediction (see text).
}
\label{fig:FIGURE3}
\end{figure}
%
%
%%%%%%%%%%%%

The 1D transport equation for a dye solute of charge $+e$ follows from Eq.\,\eqref{diff_conv}: $u c - D c \, \partial_x \phi - D \partial_x c = J_0$, with $J_0$ the solute current, independent of the $x$ location. Such equation is readily solved in each five regions of the system and assuming concentration continuity between regions, the solution for the dye solute concentration is obtained analytically. In the limit where the reservoirs are large enough to avoid finite-size effects, i.e. $(L_R/L)\mathrm{Pe}\gg1$, the concentration inside the nanochannel takes a simple expression
\begin{equation}
c \approx c_0 + c_0 \left( \mathrm e^{-\phi_D} - 1 \right) \mathrm e^{-u L / (2D)} \mathrm e^{u x / D}.
\label{eq:c3}
\end{equation}

Following the concentration normalization introduced previously for experimental data, the theoretical prediction eventually reads
\begin{equation}
\frac{c(x)-c_0}{c_m-c_0} =  \mathrm e^{- \alpha_{1D}(x) \text{Pe}},
\label{eq:c3bis}
\end{equation}
where the prefactor $\alpha_{1D}(x) = (1/2 - x/L)$ accounts for the measurement location, and where the enriched nanochannel concentration without flow $c_m$ is given by the Donnan equilibrium with reservoirs $c_m = c_0\mathrm e^{-\phi_D}$. As can be seen in Fig.\,\ref{fig:FIGURE3}\,A (middle row), this theoretical prediction Eq. \eqref{eq:c3bis} is in perfect agreement with a finite element calculation of the full problem (see Supp. Mat.) in the limit of strong Debye layer overlap $2\lambda_D/h = 10$, for which the 1D approximation is expected to stand.

\begin{figure}
\includegraphics[width=1.0\linewidth]{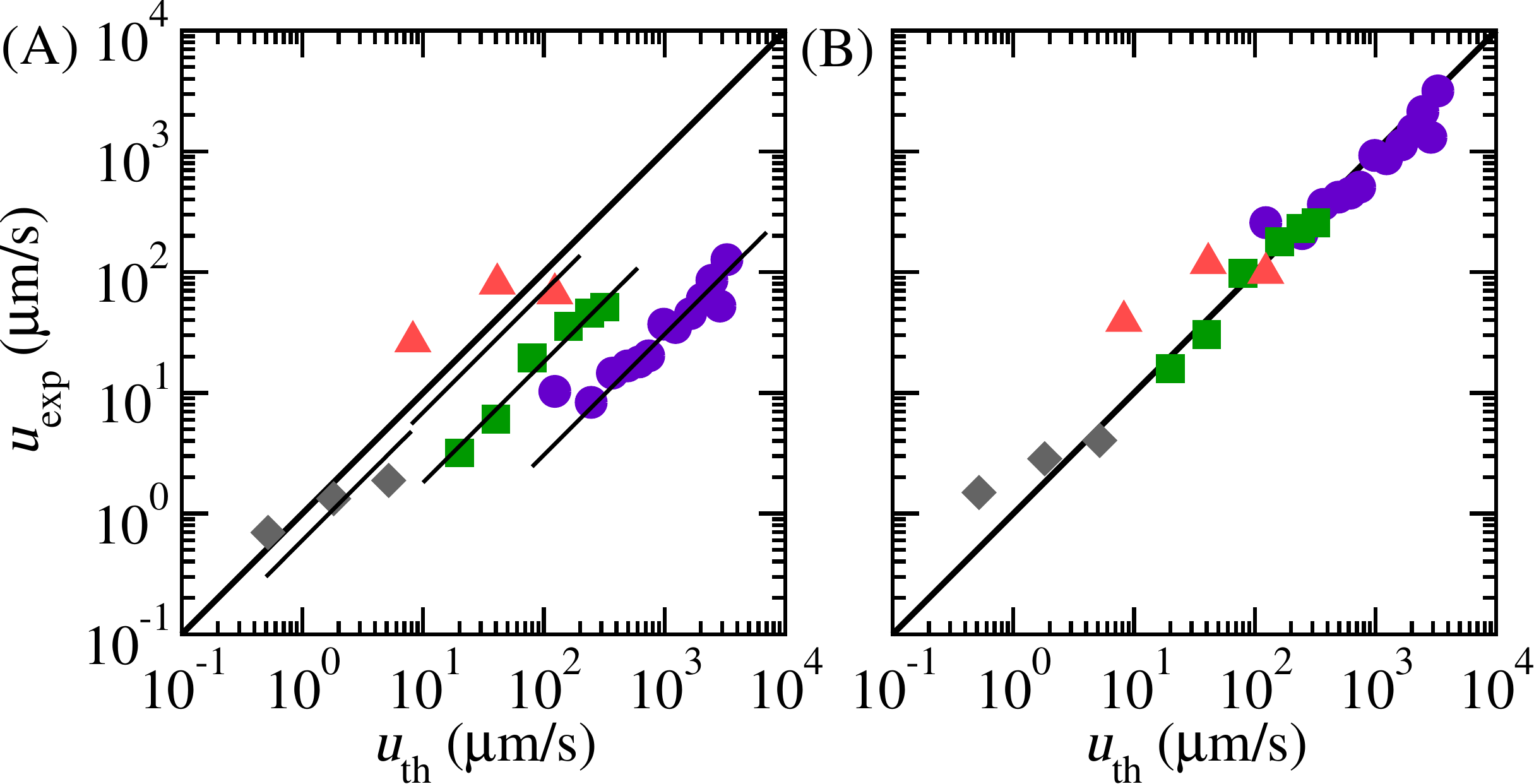}
\caption{%
Experimental velocities $u_\text{exp}$ from flow-induced concentration polarization measurements against Poiseuille-Hagen expectation $u_\text{th}$.
\textbf{(A)} Experimental velocities extracted from 1D theory Eq. \eqref{eq:c3bis}. $L=500\,$\si{\micro}m: $c_s = 0.1$\,mM ({\color{gray}$\blacklozenge$}); $L=25\,$\si{\micro}m: $c_s = 0.1$\,mM; ({\color{red}$\blacktriangle$}), $c_s = 1$\,mM ({\color{greenSimon}$\blacksquare$}), $c_s = 10$\,mM ({\color{violet}$\bullet$}). Full lines indicate $u_\text{exp.}\propto u_\text{th.}$ scaling.
\textbf{(B)} Same as (A) with experimental velocities extracted from 2D theory Eq. \eqref{eq:caveraged2}.}
\label{fig:FIGURE4}
\end{figure}

Coming back to experiments, Eq.\,\eqref{eq:c3bis} predicts the observed exponential decay with the Peclet number shown in Fig. \ref{fig:FIGURE1} and \ref{fig:FIGURE2}, and probed by varying either the driving pressure or the channel length. In addition, Eq.\,\eqref{eq:c3bis} accounts for the influence of the measuring location involved in Fig.\,\ref{fig:FIGURE2}\,A. Moreover, for the lowest salt concentration $c_s=0.1\,$mM for which the 1D assumption is approached ($2\lambda_D / h \sim 0.5$), Eq.\,\eqref{eq:c3bis} satisfactorily describes experimental data with no free parameter (Fig.\,\ref{fig:FIGURE4}\,A).

%%%%%%%%%%%%%%%%%%%%%%%%%%%%%
% Beyond 1D, effect of EDLs 
%%%%%%%%%%%%%%%%%%%%%%%%%%%%%

Despite capturing the global form of the flow-induced solute concentration polarization inside the nanochannel, the previous theoretical description fails to encompass the effect of increasing salt concentration, thus pointing to a limitation of the 1D assumption, see Fig.\,\ref{fig:FIGURE3}\,A (lower row) and Fig.\,\ref{fig:FIGURE4}\,A. In order to capture this effect, inhomogeneous distributions of all physical quantities across the nanochannel height $z$ need to be incorporated in the model. Starting from Eq. \eqref{diff_conv}, with $c$, $u$, $\phi$ and $J_0$ now including a $z$-dependency, we followed a perturbation approach of the 2D problem looking for first order solution $c \simeq c^{(0)} + \text{Pe} \, c^{(1)}$ of a Peclet development, in the limit of $\text{Pe} \ll 1$.

Such approach provides the $z$-averaged concentration inside the nanochannel
\begin{equation}
\left< c \right>_z = c_0 \left< \mathrm e^{-\phi} \right>_z + c_0 \text{Pe} \left( \dfrac{\left< u \mathrm e^{-\phi} \right>_z}{\left< u \right>_z} - 1 \right) \dfrac{x}{L},
\label{eq:caveraged}
\end{equation}
where the Peclet number is now defined according to the z-averaged flow profile $\langle u\rangle_z$ and where $\phi(z)$ is the $x$-invariant reduced electric potential without flow (see Supp. Mat. for details). Proceeding similarly for the 1D problem we find that the 2D solution can be matched onto the 1D expression providing $\mathrm{Pe}$ is replaced by $\alpha_{2D}\mathrm{Pe}$, with
\begin{equation}
\alpha_{2D}=\dfrac{\left< (c - c_0) u \right>_z}{\left< c - c_0 \right>_z \left< u \right>_z},
\label{eq:alpha2D}
\end{equation}
where $c_0$ is the reservoir concentration,
and where we have used for the Donnan potential $\mathrm e^{-\phi_D} = \left< \mathrm e^{-\phi} \right>_z$.

For larger Peclet number, we thus propose an ansatz solution which accounts for 2D effects in non-overlapping confinements, and that is based on the 1D Eq. \eqref{eq:c3bis}:
\begin{equation}
\frac{\langle c\rangle_z - c_0}{\langle c_m\rangle_z - c_0}= \mathrm e^{-\alpha_{1D}(x) \alpha_{2D} \text{Pe}}.
\label{eq:caveraged2}
\end{equation}
For the mapping factor $\alpha_{2D}$, a simple estimate can be obtained by assuming weak surface potentials together with the additivity of opposite-wall contributions. From Eq. \eqref{eq:alpha2D} and with $u(z) = (z^2 - (h/2)^2 ) \Delta p/ ( 2 L \eta )$, we obtain $\alpha_{2D}\simeq 3 [ \tilde h \coth ( \tilde h ) - 1 ] / \tilde h^2$ that only depends on the reduced thickness $\tilde h = h / ( 2 \lambda_D )$.

As can be seen in Fig.\,\ref{fig:FIGURE3}, where our ansatz Eq. \eqref{eq:caveraged2} is compared to finite elements results, the exponential decay in $\mathrm{Pe}$ is perfectly recovered whatever the overlapping parameter. Moreover, the simple estimate for the mapping factor $\alpha_{2D}$ well captures the effect of Debye layer confinement. Note however a moderate shift between  analytical prediction and finite element results for $2 \lambda_D / h < 1$ (Fig.\,\ref{fig:FIGURE3}\,C). This deviation may result from some of the effects neglected in the present treatment such as: couplings between $x$ and $z$ directions, which are likely to occur at the channel entrances but were neglected in the theoretical calculations;  a different flow field in the reservoirs as compared with the model assumption due to the large reservoirs height; etc.

%%%%%%%%%%%%
%
%
%
%
%%%%%%%%%%%%

%%%%%%%%%%%%%%
% Conclusion %
%%%%%%%%%%%%%%

Combining the theoretical description given by Eq.\eqref{eq:caveraged2}, together with $\alpha_{2D}$ as estimated from finite element calculations, we are now able to extract an experimental flow velocity $u_\mathrm{exp}$ from the measured flow-induced concentration polarization presented in Fig.\,\ref{fig:FIGURE2}. Comparing with the expected velocity $u_\text{th}$, as estimated from the Poiseuille-Hagen relationship for known imposed pressures, we obtain an excellent agreement evidenced in Fig.\,\ref{fig:FIGURE4}\,B. As mentioned, accounting for 2D effects across the channel height is essential for reliable flow measurements (Fig.\,\ref{fig:FIGURE4}\,A) 

In terms of flow rate sensitivity the present results reach sub-fL/s limits with $Q\simeq0.4\,$fL/s, making of the flow-induced shift in Donnan equilibrium an ultra-sensitive probe of mass transport in single nanochannels. Note that because effects actually depends on the flow velocity, reducing the channel width down to the focused spot size would push resolution down a decade without modifying the signal to noise ratio. Similarly, thought at as an integrated sensor in series with other nanofluidic systems, the system sensitivity can be further increased by varying the nanoslit length. Based on ubiquitous concentration polarization effects, its minimal probe requirements and sensitivity to transport processes, we believe that this work can contribute to the fast development of nanofluidics both for fundamental aspects and towards high impact applications.

\section*{ACKNOWLEDGMENTS}
\begin{acknowledgments}
The authors thank R. Fulcrand, C. Lee and P. Joseph for Chips nanofabrication; L. Joly, C. Cottin-Bizonne and A.-L. Biance for fruitful discussions; L. Joly and M. Stojanova for comments on the manuscripts. S. G. acknowledge financial support from L. Bocquet through the European Research Council program Micromegas project.
\end{acknowledgments}

%
%%%%%%%%%%%%%%%%%%%
% Biblio
%%%%%%%%%%%%%%%%%%%
\bibliographystyle{ieeetr} % uncomment to show titles
\bibliography{FCS_manuscript}
\end{document}